\documentclass[aps,pre,twocolumn,showpacs,floatfix,superscriptaddress]{revtex4-1}
\usepackage{epsfig}
\usepackage{amsmath}

\begin{document}

\title{Brittle to Ductile Transition in a Fiber Bundle with Strong Heterogeneity}

\author{Korn\'el Kov\'acs}
\affiliation{Department of Theoretical Physics, University of
Debrecen, P. O. Box: 5, H-4010 Debrecen, Hungary}

\author{Raul Cruz Hidalgo}
\affiliation{Departamento de F\i sica, Facultad de Ciencias, Universidad de Navarra, 
31080 Pamplona, Spain}

\author{Ignacio Pagonabarraga}
\affiliation{Departament de F\'{\i}sica Fonamental, Universitat de
Barcelona, Carrer Mart\'{\i} i Franqu\'es 1, E-08028 Barcelona, Spain}

\author{Ferenc Kun}\email{Email address:ferenc.kun@science.unideb.hu}
\affiliation{Department of Theoretical Physics, University of
Debrecen, P. O. Box: 5, H-4010 Debrecen, Hungary}

\date{\today}
             
\begin{abstract} 
We analyze the failure process of a two-component system with widely
different fracture strength in the framework of a fiber bundle model
with localized load sharing. A  fraction $0\leq \alpha \leq 1$ 
of the bundle is strong and it is represented by unbreakable fibers, while fibers of the weak component 
have randomly distributed failure strength. Computer simulations revealed that 
there exists a critical composition $\alpha_c$ 
which separates two qualitatively 
different behaviors: below the critical point the failure of the bundle is 
brittle characterized by 
an abrupt damage growth within the breakable part of the system.
Above $\alpha_c$, however, the macroscopic response becomes ductile providing 
stability during the entire breaking process. 
The transition occurs at an astonishingly low fraction of strong fibers
which can have importance for applications.  
We show that in the ductile phase the size distribution of breaking
bursts has a power law functional form with an exponent $\mu=2$ followed by an
exponential cutoff. In the brittle phase the power law also prevails but
with a higher exponent $\mu=9/2$. The transition between the two phases 
shows analogies to continuous phase transitions. 
Analyzing the microstructure of the damage, it was found that at the 
beginning of the fracture process cracks nucleate randomly, while later on
growth and coalescence of cracks dominate which give rise to power law
distributed crack sizes.

\end{abstract}
\pacs{05.90+m, 81.40.Np, 64.60.A-}
\maketitle

\section{Introduction}
Beyond its engineering relevance, the complexity of the fracture of brittle and 
ductile materials is of great interest for physics and materials science, as well.
Recently, the application of statistical physics has revealed
interesting novel aspects of the damage and fracture of heterogeneous
materials increasing our understanding both on the micro
and macro scales of fracture processes \cite{alava_statistical_2006}. 
Most of these theoretical investigations relay on mesoscopic discrete models
such as fiber bundles 
\cite{zapperi_plasticity_1997,pradhan_failure_2010,pradhan_crossover_2005,hidalgo_bursts_2001,hidalgo_fracture_2002} 
and lattices of fuses \cite{batrouni_fracture_1998,nukala_percolation_2004}, 
springs, or beams \cite{daddetta_application_2002,raischel_simple_2005,
nukala_fracture_2010}, where
disorder is typically captured by the random strength of elements. 
Analytic calculations and
computer simulations have revealed that for a broad class of disorder
distributions the fracture of heterogeneous materials exhibit
universal aspects both on the micro- and macro levels: the size of
bursts has a power law distribution with universal exponents
\cite{hidalgo_avalanche_2009,kloster_burst_1997,pradhan_crossover_2005,hidalgo_bursts_2001,hidalgo_fracture_2002},
furthermore, macroscopic failure occurs in the form of localization
after a precursory sequence of microcracking \cite{kun_universality_2008,
saichev_andrade_2005,andersen_tricritical_1997}.
Interesting analogies have been established between the failure of
materials and phase transitions and critical phenomena 
\cite{andersen_tricritical_1997,zapperi_plasticity_1997,zapperi_first-order_1997,moreno_fracture_2000,
kovacs_critical_2008}.

However, real materials are often of composite nature, {\it i.e.} they
have two or more components with widely different properties which
provide an improved fracture toughness. For instance, in fiber
reinforced composites where fibers are embedded in a carrier matrix,
the matrix material typically has much lower fracture
strength than the fibers. This widely different strength of the
components combined with appropriate coupling 
results in an improved damage tolerance which has a high relevance for 
applications \cite{matthews_composite_1994,phoenix_proc_roy_soc_1983}. 
It has been shown that 
in heterogeneous materials 
varying the local mechanical response and of the amount of disorder of the 
components one can achieve a transition from brittle 
to quasi-brittle or even to ductile failure 
\cite{amitrano_brittle-ductile_2003,picallo_brittle_2010,
raischel_failure_2006,krajcinovic_damage_1996}. 

In this paper we study the effect of strong heterogeneity on the
fracture process of disordered materials based on a fiber bundle model (FBM)
with localized interaction. Our model is composed
of two subsets of fibers with widely different fracture behavior:
fibers of one of the subsets are strong in the sense that they can
support any load and never break, while fibers of the other type
are weak characterized by a probability distribution of failure
strength. The two components form a homogeneous mixture on a square lattice 
with periodic boundary conditions in both directions. It is a crucial
element of our approach that after failure events the load of broken fibers
is equally redistributed solely over their intact nearest neighbors 
leading to high stress concentration around failed regions.
Such localized load sharing systems are known to be very brittle, however,
our investigations demonstrate that the presence of strong fibers moderates
the effect of stress inhomogeneities and leads to a ductile macroscopic response
when exceeding a critical fraction. 
Ductility is qualified in our system by the stable growth of cracks at the microscopic level
which leads to the emergence of a relatively long plateau regime in the macroscopic 
response of the system.
We explore by computer simulations how the 
brittle-ductile transition occurs on the micro-scale by investigating the size 
distribution of bursts and the micro-structure of damage.

\section{Mixture of weak and strong fibers}

We consider a parallel bundle of fibers organized on a
square lattice of size $L$. Under an increasing external load the fibers present a
linearly elastic behavior characterized by the same Young modulus $E$. 
To model the two-component mixture, a fraction $0\leq \alpha \leq 1$ of the $N=L^2$ 
fibers of the lattice is
considered to be unbreakable, {\it i.e.\ } they can support any load without
failure.
These strong fibers of number $N_s=\alpha\cdot N$ are distributed randomly
all over the lattice without
any spatial correlations. The remaining $1-\alpha$ fraction of fibers of number
$N_w=(1-\alpha)N$  are considered to be weak, i.e. they have only a
finite load bearing capacity and break when the 
local load on them exceeds their failure threshold. The breaking thresholds of
weak fibers $\sigma_{th}^i\ \  i=1,\ldots,N_w$ are independent, identically
distributed random variables
with a probability density $p(\sigma_{th})$ and distribution function
$P(\sigma_{th})$. For simplicity, we consider a uniform distribution between
$0$ and $1$ so that $p(\sigma_{th})$ and $P(\sigma_{th})$ take the forms
$p(\sigma_{th})=1$ and $P(\sigma_{th})=\sigma_{th}$, respectively.
The bundle is subject to a quasi-statically increasing external load $\sigma$ parallel
to the direction of fibers.
When a fiber breaks in the bundle its load has to be shared by the remaining 
intact fibers. As a crucial element of the model,  we assume that the load sharing is completely 
localized in the  system so that the excess load after failure events is equally redistributed over 
the intact nearest  neighbors of failed fibers. This localized load sharing (LLS) introduces 
spatial correlations in the system which makes it impossible to carry out
analytical calculations. 

\subsection{Equal Load Sharing (ELS)}

In a previous work \cite{hidalgo_universality_2008}, we have shown that in the limit case of equal load sharing (ELS)
the most important characteristics of the fracture process of the two component system
can be obtained in closed analytical form. Under ELS conditions the constitutive 
relation $\sigma^{\sc ELS}(\varepsilon)$ of the model can be written as
\begin{equation}
\sigma^{\sc ELS}(\varepsilon) =  (1-\alpha) \left[1-P(E\varepsilon)\right] E\varepsilon + \alpha E\varepsilon,
\label{const1}
\end{equation}
where $\varepsilon$ denotes the strain of the system. The
first term of Eq.\ (\ref{const1}) 
accounts for the load bearing capacity of the surviving fraction
of {\it weak} elements, and the second one represents the stress
carried by the {\it unbreakable} subset of the system. 
This constitutive equation $\sigma^{\sc ELS}(\varepsilon)$ is illustrated in  Fig.\
\ref{fig:constitutive} as continuous curves for uniformly distributed failure thresholds
at several values of $\alpha$.
Note, that one recovers the usual FBM constitutive behavior \cite{kloster_burst_1997} in the
limiting case of $\alpha=0$, when the bundle is only composed of weak
fibers.  Those solutions usually present a parabolic maximum, which  
defines the critical deformation $\varepsilon_c^{\sc ELS}$  
and strength $\sigma_c^{\sc ELS}$ under stress-controlled conditions \cite{hidalgo_universality_2008}. 
For finite values of $\alpha$, all the weak fibers break for large enough $\varepsilon$  so
that the first term of Eq.\ (\ref{const1}) goes to zero
while the {\it unbreakable} fibers overtake the entire external load. 
Consequently, the constitutive curves 
in Fig.\ \ref{fig:constitutive} tend asymptotically to straight lines with slope
$\alpha E$.  It can be seen in Fig.\ \ref{fig:constitutive} that for low values of $\alpha$ the local
maximum of $\sigma^{\sc ELS}(\varepsilon)$ prevails but its position
$\varepsilon_c^{\sc ELS}$ and value  $\sigma_c^{\sc ELS}$ are monotonically
increasing with $\alpha$ \cite{hidalgo_universality_2008}. 

For the case of an uniform distribution, we have analytically found that
the position of the maximum depends on $\alpha$ following $\sigma_c^{\sc ELS}(\alpha)=1/2(1-\alpha)$ 
\cite{hidalgo_universality_2008}. 
This holds for $\alpha\le\alpha_c^{\sc ELS}$ with the critical value of the control parameter
$\alpha_c^{\sc ELS}=1/2$. At $\alpha_c^{\sc ELS}$ the value of $\varepsilon^{\sc ELS}_c$ coincides with
the upper bound of strength values $\sigma_{th}^{max}/E=1$. The
parabolic shape of the constitutive curve prevails even for
$\alpha>\alpha_c^{\sc ELS}$ but $\sigma^{\sc ELS}(\varepsilon)$ becomes linear at
$\varepsilon=\sigma_{th}^{max}/E$ before reaching the maximum, so that the
rest of the parabola cannot be realized.  Complementary, it was also shown analytically 
that the existence of this critical point and the qualitatively different 
shape of the constitutive curve below and above $\alpha_c^{\sc ELS}$  
have a substantial effect on the microscopic breaking of the system 
\cite{hidalgo_universality_2008}. 


\subsection{Local Load Sharing (LLS)}

In this work, we examine the case of localized load redistribution.
In the following we show that the presence of unbreakable fibers results 
in a broad spectrum of novel behaviors when the stress field is inhomogeneous. 
In the two-component mixture, the unbreakable fibers act as a 
load reservoir, i.e.\ the load they carry, especially the load increments they receive
from their broken weak neighbors, does not contribute to the breaking process;
from the viewpoint of breakable fibers this load is dissipated. It has the
consequence that increasing the fraction of strong fibers reduces the stress
concentration around failed regions. 
The value of $\alpha$ controls the relevance of the stress concentration and the disorder during the failure process.
Thus, it has interesting consequences on both the micro and macro-scale response of the system.

In order to obtain a detailed understanding of the effect of stress localization 
on the fracture process in two-component mixtures, we carried out computer 
simulations on a square lattice of size $L= 401$ with periodic boundary conditions
in both directions varying the value of $\alpha$ between 0 and 1. 
Averages were calculated over 100 samples with different realizations 
of disorder.

\begin{figure}
\begin{center}
\epsfig{bbllx=5,bblly=512,bburx=350,bbury=790,file=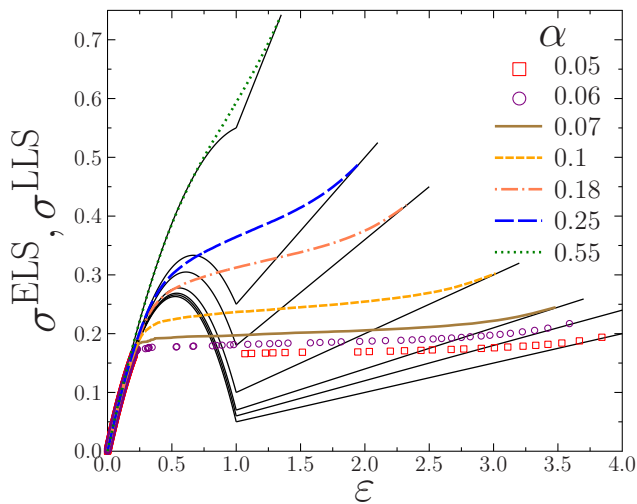,
  width=8.5cm}
   \caption{{\it (Color online)} The constitutive behavior of
   the two-component bundle for ELS (black continuous lines) and $LLS$ (dashed-dotted lines with 
   different colors). The transition from brittle to ductile behavior can clearly 
   be observed as $\alpha$ increases from bottom to top.
\label{fig:constitutive}}
  \end{center}
\end{figure}

Figure \ref{fig:constitutive} illustrates the constitutive
curve of the $LLS$ bundle $\sigma^{\sc LLS}(\varepsilon)$ for
several values of $\alpha$ in comparison with the corresponding 
ELS results $\sigma^{\sc ELS}(\varepsilon)$. 
At $\alpha=0$ when the system contains only breakable fibers we
recover the former $LLS$ results, namely, the $LLS$ constitutive curve
follows the ELS analytic solution \cite{hansen_burst_1994,kloster_burst_1997,
kun_damage_2000-1,pradhan_failure_2010,newman_time-dependent_2001,
phoenix_time-dependent_2009,raischel_local_2006}. However, the critical stress
$\sigma_c^{LLS}$ and strain $\varepsilon_c^{LLS}$, where macroscopic
failure occurs, are significantly lower than the ELS values $\sigma_c^{ELS}$ and 
$\varepsilon_c^{ELS}$ falling in the 
linear regime of the constitutive curve. This result implies that under $LLS$
conditions FBMs exhibit a more brittle behavior than for long range load redistribution.
Under stress controlled loading macroscopic failure occurs
in the form of a catastrophic avalanche at $\sigma_c^{LLS}$ during which all remaining 
fibers break. For finite $\alpha$ values it can be observed in Fig.\ 
\ref{fig:constitutive} that the
structure of $\sigma^{LLS}(\varepsilon)$ has two substantially 
different regimes: increasing the external
load $\sigma$ for $\alpha$ values close to zero the constitutive curve has a
finite horizontal jump then it continues again along the ELS analytic solution.
The discontinuity of $\sigma^{LLS}(\varepsilon)$ is the consequence of a major
avalanche in which a macroscopic fraction of weak fibers breaks.
However, not all the weak fibers do break at this large event.
The smooth convergence of $\sigma^{LLS}(\varepsilon)$ to the asymptotic linear curve
indicates that a small fraction of fibers survives and gradually breaks as the
external load $\sigma$ is further increased. It is important to emphasize that
there exists a well defined value $\alpha_c$ of the
control parameter at which the jump and the dominating avalanche disappears and 
$\sigma^{LLS}(\varepsilon)$ becomes a continuous monotonically increasing 
function. Increasing $\alpha$ above $\alpha_c$ the qualitative form of the
constitutive curve does not change anymore.
It is a very important feature of the system
that in the regime $\alpha > \alpha_c$ the macroscopic response exhibits 
ductile behavior, i.e.\ instead of the abrupt damage growth
observed below
the critical point, the fracture process retains stability up to very large strain
values.This brittle-ductile transition occurs at an astonishingly low fraction 
of strong fibers $\alpha_c^{LLS}\approx 0.059(5)$ compared to the corresponding
mean-field counterpart $\alpha_c^{LLS}\ll \alpha_c^{ELS}$.

In local load sharing approximation, the critical strength $\sigma_c$ of an homogeneous system of breakable fibers ($\alpha=0$) 
depends logarithmically on the system size. Consequently, $\sigma_c$ vanishes  in the thermodynamic limits. 
However, for $\alpha>0$, when increasing the system size, the weight of both components (breakable and unbreakable)
increases with the same proportion. Furthermore, in the thermodynamic limits 
one could expect a similar logarithmic convergence but to a finite value $\sigma_c(\alpha)$.
Therefore, we expect that the system size dependence of $\sigma_c(\alpha)$ won't significantly 
modify the shape of the curves shown in Fig.\ref{fig:constitutive}, as well as the value of 
$\alpha_c^{LLS}$.

The unique macroscopic response of the system is the fingerprint of the special
features of the microscopic fracture process. Due to the localized interaction
of fibers, both spatial and temporal correlations arise at the micro-level: 
 breaking fibers trigger bursts which are spatially localized
and result in correlated growth of cracks.
In order to characterize the breaking process on the micro scale, we analyze the
statistics of bursts of simultaneously failing fibers and the spatial structure
of broken clusters. 

\section{Burst size distribution}

The failure process of $LLS$ fiber
bundles has recently been explored in detail by computer simulations in the
case of $\alpha=0$ where the bundle is only composed of weak fibers
\cite{hansen_burst_1994,kloster_burst_1997,kun_damage_2000-1,pradhan_failure_2010,newman_time-dependent_2001,
phoenix_time-dependent_2009,raischel_local_2006, phoenix_proc_roy_soc_1983}. 
It was found that under an increasing external load first the weakest fibers break 
randomly and homogeneously over the entire system.
As the external load increases the load dropped by the broken fibers becomes
sufficient to give rise to additional breakings. As a result, breaking fibers can 
trigger avalanches of breaking events which are spatially correlated giving rise to
growing clusters of broken elements. Connected clusters of broken fibers on the lattice
can be interpreted as cracks in FBMs \cite{kun_damage_2000-1,phoenix_proc_roy_soc_1983}.
The microscopic origin of the brittle behavior, observed in the limit of $\alpha=0$ 
in Fig.\ \ref{fig:constitutive},
is that along the perimeter of growing cracks a high load
is concentrated, which easily initiates a catastrophic avalanche 
already at crack sizes much smaller than the system size. 
\begin{figure}
\begin{center}
\epsfig{bbllx=5,bblly=510,bburx=345,bbury=790,file=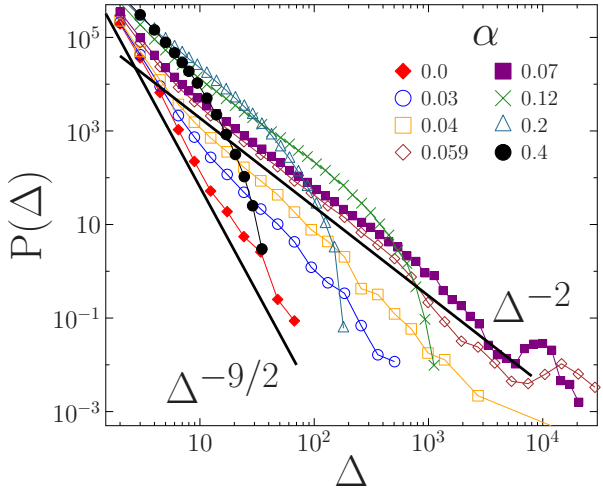,
  width=8.0cm}
   \caption{{\it (Color online)} Avalanche size distributions $P(\Delta)$ at
   different values of $\alpha$ below and above $\alpha_c$.
Straight lines with slope $9/2$ and $2.0$ are drawn to guide the eye.
The value of $\alpha$ increases from bottom to top in the range $\Delta < 10$.} 
\label{fig:fig2}
  \end{center}
\end{figure}
As a consequence, in our simulations the completely weak system cannot tolerate large bursts
and the avalanche size distribution $P(\Delta)$ becomes a rapidly decreasing function.
The data analysis shows that the distribution function is a power
law with a relatively large exponent (see Fig.\ 
\ref{fig:fig2})
\begin{equation} 
P(\Delta)\sim\Delta^{-\mu}, \ \ \ \ \ \ \mbox{where} \ \ \ \ \ \  \mu\approx 9/2,
\end{equation}
in agreement with former predictions 
\cite{hansen_burst_1994,raischel_local_2006,kun_damage_2000-1}.
For increasing $\alpha$ simulations revealed that the reduction of stress  
concentration by strong fibers allows the system for larger avalanches. 
It can be observed in Fig.\ \ref{fig:fig2} that the
exponent $\mu\approx 9/2$ still remains for small avalanche sizes, while for large
avalanches a crossover occurs to a lower exponent $\mu\approx 2.0$. 
Approaching the critical value $\alpha_c$, the second power
law regime spans over nearly three orders of magnitude in a system of size
$L=401$. Note the bump of $P(\Delta)$ for the
largest avalanches. Since the largest avalanche of the system is always followed
by a few small ones, no catastrophic event can be
identified. 
Consequently, all the avalanches were taken into account when performing the counting.
Hence, the statistics of the largest avalanches forms a bump of Gaussian shape 
on the total distribution.
Fig.\ \ref{fig:fig2} shows that above $\alpha_c$ the bump in $P(\Delta)$, associated to large avalanches, 
disappears and  an exponential cutoff develops instead.
Increasing $\alpha$ further does not affect the power law distribution, 
although the cutoff avalanche size decreases.

\begin{figure}
\begin{center}
\epsfig{bbllx=10,bblly=475,bburx=350,bbury=770, 
file=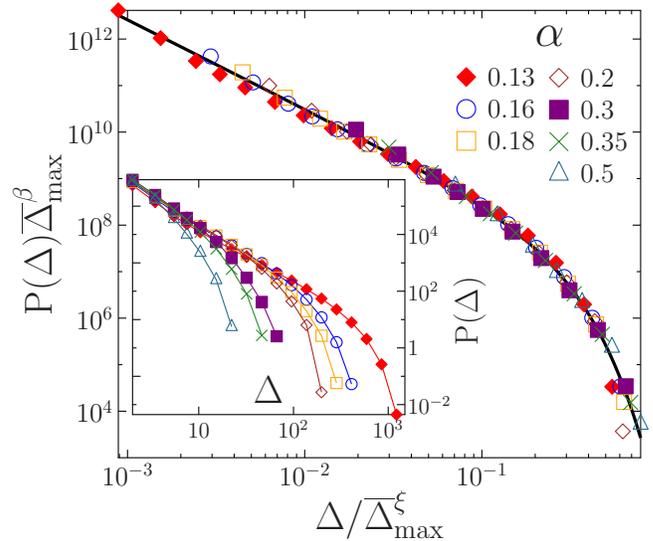, width=8.5cm}
   \caption{{\it (Color online)} Scaling plot of avalanche size distributions above
$\alpha_c$ using the average size of the largest avalanche
$\overline{\Delta}_{max}$ as scaling variable. An excellent collapse
is achieved with the exponents $\xi=1.24$ and $\beta=2.5$. 
The bold line represents the fit with Eq.\ (\ref{eq:scalfunc}), where
$\mu=1.97$ was obtained. The original distributions $P(\Delta)$ are
presented in the inset where $\alpha$ increases from right to left.}  
\label{fig:scaled_above}
  \end{center}
\end{figure}

Scaling analysis revealed that rescaling the two axis by appropriate
powers of a characteristic avalanche size, the
distributions $P(\Delta)$ above the critical point $\alpha_c$ can be
collapsed onto a master curve.
Using the average size of the largest avalanche $\overline \Delta_{max}$ as
scaling variable, an excellent quality data  collapse is obtained in Fig.\
\ref{fig:scaled_above} which implies the scaling structure of the burst size 
distribution
\begin{equation}
P(\Delta)=\overline{\Delta}_{max}^{-\beta}f(\Delta/\overline\Delta_{max}^{\xi}),\ \
\ \ \mbox{for}\ \ \ \ \alpha >\alpha_c.
\end{equation}
The best collapse is achieved for the exponents $\xi=1.24$ and $\beta=2.5$
as seen in Fig.\ \ref{fig:scaled_above}.
It is important to emphasize that the scaling
function $f$ has the functional form
\begin{eqnarray}
f(\Delta/\overline\Delta_{max}^{\xi}) \sim \Delta^{-\mu}\exp{(-\Delta/\overline
\Delta_{max}^{\xi})}, 
\label{eq:scalfunc}
\end{eqnarray}
i.e., it can be described as a power law followed by an exponential cutoff. The value
of the exponent $\mu$ has been obtained by fitting as $\mu=2.0\pm 0.05$. Note that due
to the condition of normalization the three exponents $\xi$, $\beta$, and $\mu$
must fulfill the scaling relation $\beta=\xi\mu$. Substituting the numerical values
good agreement can be found. The above scaling analysis also demonstrates that the 
value of the exponent $\mu$ of the burst size distribution is constant above the
critical point so that any apparent change of $\mu$ in Fig.\ \ref{fig:scaled_above}
can be attributed to the moving cutoff. 

\begin{figure}
\begin{center}
\epsfig{bbllx=10,bblly=500,bburx=343,bbury=785,file=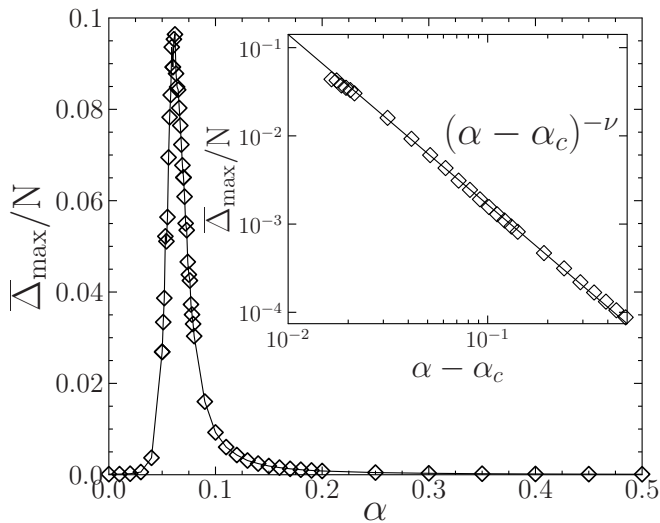,
  width=8.5cm}
   \caption{ Average size of the largest avalanche
   $\overline{\Delta}_{max}$ as a function of $\alpha$. The position of the sharp peak
   $\alpha_c$ coincides with the critical value of $\alpha$ we identified based
   on the constitutive curves. Inset: the same quantity plotted as a function
   of the distance from the critical point $\alpha-\alpha_c$ for $\alpha>\alpha_c$. 
A straight line of
   slope $1.95$ is drawn to guide the eye.}
\label{fig:avalmax}
  \end{center}
\end{figure}

Computer simulations have shown that the characteristic avalanche size 
$\overline\Delta_{max}$ as a
function of $\alpha$ has a sharp peak (see Fig.\ \ref{fig:avalmax}) at the same $\alpha_c$  
which has been defined based on 
the constitutive curve of the system in Fig. \ref{fig:constitutive}. Replotting
$\overline\Delta_{max}$ as a function of $\alpha-\alpha_c$ in the inset of Fig.\ 
\ref{fig:avalmax} a straight line has been obtained on a double logarithmic plot which
implies the power law form
\begin{equation}
\overline\Delta_{max} \sim \left( \alpha-\alpha_c \right)^{- \nu}, \ \ \ \ \mbox{
for } \ \ \ \ \alpha > \alpha_c.
\end{equation}
The value of the correlation length exponent $\nu$ was obtained numerically as
$\nu=1.95\pm 0.07$.
The result shows that approaching the critical point $\alpha_c$
from above the characteristic avalanche size has a power law divergence. 
When $\alpha$ is increased starting from zero, during the damaging of the bundle
an increasing fraction of load is carried by the strong fibers.
This induces a screening effect and part of the system becomes inaccessible for the damage growth.
Consequently, the active load decreases and the bundle can survive larger avalanches, 
which reach the scale of the system size at $\alpha=\alpha_c$.

Comparing to the mean field solution of the model \cite{hidalgo_universality_2008},
it is noticeable that the burst size exponent $\mu$ of the regime $\alpha\geq \alpha_c$
is smaller, while the correlation length exponent $\nu$ is larger when the stress
redistribution is localized. In Ref.\ \cite{hidalgo_avalanche_2009} we have shown
by analytical means that the burst size exponent of FBMs takes the value $\mu=2$
when the constitutive curve has a long plateau preceded by an increasing regime
with only a slight non-linearity. Our numerical results obtained for two-component
mixtures is in perfect agreement with the analytical predictions of Ref.\
\cite{hidalgo_avalanche_2009}.

\section{Micro-structure of damage}
Contrary to equal load sharing, under $LLS$ conditions 
fibers breaking in a burst form a connected set which may be part of
a more extended cluster of broken fibers generated by previous avalanches. 
Therefore, the cluster structure and the avalanches of breaking fibers 
become correlated. In the following we explore how the micro-structure
of damage evolves under an increasing external load $\sigma$ for different 
compositions of the system and establish the relation of crack growth and 
breaking avalanches.

\begin{figure}
\begin{center}
\epsfig{bbllx=0,bblly=0,bburx=585,bbury=585,file=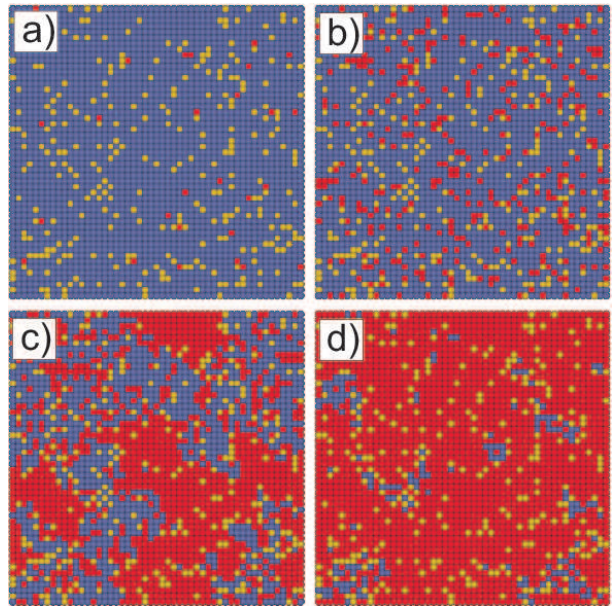,
  width=8.0cm}
  \end{center}
   \caption{{\it (Color online)} Snapshots of a damaging bundle 
   of size $L=50$ at four values of $\varepsilon$ during the loading process with $\alpha=0.07$. 
Blue (grey) and yellow (light grey) colors represent intact weak and strong fibers, respectively, 
while red (dark grey) stands for the broken ones. 
\label{fig:continous_state_matrix}} 
\end{figure}

\subsection{Nucleation, growth, and merging of cracks}
It has been discussed that in the limit of $\alpha=0$ 
all clusters of broken fibers are small
compared to the system size and they are randomly dispersed over the
entire bundle. Macroscopic failure is typically driven by the breaking
of a single fiber which is located along the boundary of a larger
cluster where the stress concentration is high 
\cite{kun_damage_2000-1,zapperi_avalanches_1999,zapperi_first-order_1997,hidalgo_fracture_2002,bosia_hierarchical_2010,
phoenix_proc_roy_soc_1983}. 
However, at finite values of
$\alpha$ the presence of strong fibers 
substantially influences the cluster structure of broken fibers of the
weak component: strong fibers decrease the load transferred to the weak
ones after breaking events which reduces the stress concentration
around failed regions. As a consequence, at low values of $\alpha$
strong fibers let the system grow larger avalanches, and hence, more
extended clusters of broken fibers can occur. However, at high 
$\alpha$, strong fibers develop a counter effect, i.e.\ the load transferred
to weak fibers after failure events will be so much reduced that it
limits the propagating bursts and growing clusters. 
Since at a given value of
$\alpha$ the bundle is a random mixture of weak and strong fibers, the underlying
structural disorder of the mixture has also an additional effect. 
For the two-component system a direct mapping can be established to percolation lattices where
$p_s=\alpha$ and $p_w=1-\alpha$ are the occupation probabilities of strong
and weak fibers, respectively, with $p_s+p_w=1$. 
In the regime $\alpha\leq p_c$, ($p_s \leq p_c$ and $p_w \geq p_c$) where $p_c$ corresponds to the critical 
occupation probability of site percolation on a square lattice \cite{stauffer_introduction_1992},
there exists a dominating cluster of weak fibers (spanning cluster).
However, for $\alpha > p_c$ the strong fibers
form a spanning cluster, while all weak clusters are
small. The intermediate regime $(1-p_c) < \alpha < p_c$ is also
interesting because here none of the components have a spanning
cluster since $p_w<p_c$ and $p_s<p_c$ hold. 

Figure \ref{fig:continous_state_matrix} presents an example for the
evolution of the 
cluster structure of the system for an increasing load $\sigma$ (for
demonstration purposes a relatively small lattice $L=50$ is considered). 
It can be seen in Fig.\ \ref{fig:continous_state_matrix}$(a)$ 
that as the external load $\sigma$ 
is increased, randomly dispersed clusters of broken fibers grow. 
In the example $\alpha=0.07$ is slightly above 
the critical point $\alpha_{c}$, hence, clusters become so large
that they can also merge (see Fig.\ \ref{fig:continous_state_matrix}$(b,c)$).
Finally, in Fig.\ \ref{fig:continous_state_matrix}$(d)$ nearly all weak fibers are broken 
so that cracks are identical with the clusters of the underlying weak component.
It has been pointed out that the probability distribution of failure thresholds, especially
the functional form of its tail may also affect the cluster structure of broken fibers
\cite{phoenix_proc_roy_soc_1983}.
For simplicity, our analysis focuses solely on the case of uniformly distributed 
thresholds, however, the results on the micros-structure of damage may slightly change
for other types of disorder.   

In order to give a quantitative characterization of the evolution of
the micro-structure of damage we determined the average size $\left<S_{av}\right>$
and the average number $\left<n_c\right>$ of clusters of broken fibers
as a function of the deformation $\varepsilon$ during the loading process.
The average cluster size $S_{av}$ is defined as the ratio
of the second and first moments of cluster sizes $S_i$
\begin{eqnarray}
S_{av} = \frac{\sum_i S_i^2}{\sum_i S_i},
\end{eqnarray}
\begin{figure}
\begin{center}
\epsfig{bbllx=5,bblly=255,bburx=350,bbury=775,
file=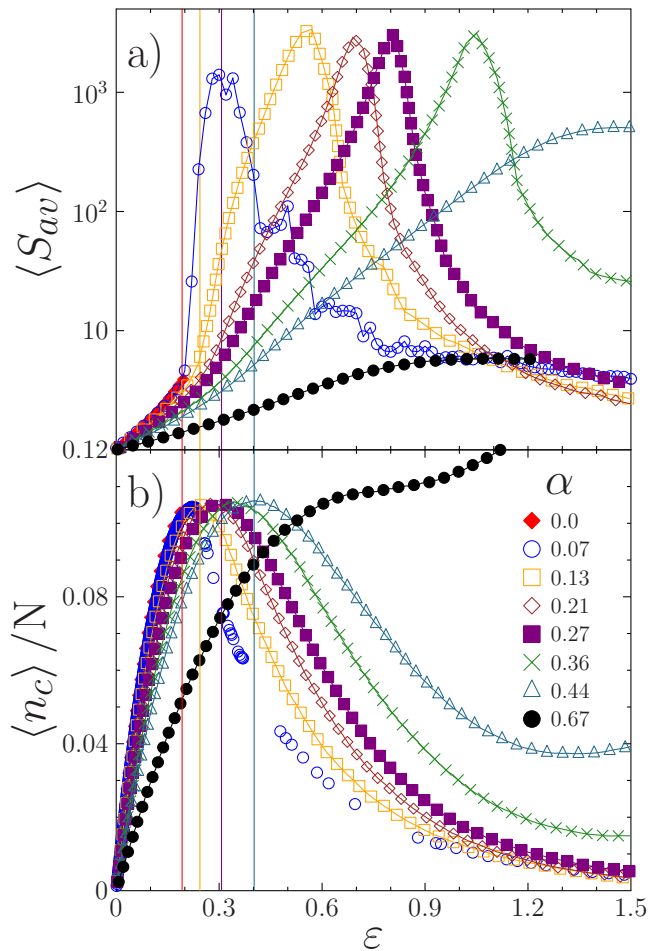, width=8.5cm}
  \end{center}
   \caption{{\it (Color online)} Average cluster size $\left<S_{av}\right>$ $(a)$ 
and average number of clusters  $\left<n_c\right>$
$(b)$ as function of $\varepsilon$ for several compositions $\alpha$ of the system.
For clarity, the vertical straight lines indicate the position $\varepsilon_m^{nc}$ of the maximum 
of $\left<n_c\right>$ for a few $\alpha$ values: $0.0, 0.13, 0.27, 0.44$.} 
\label{fig:clust_num}
\end{figure}
where the largest cluster is always omitted in the summation. 
Then $\left<S_{av}\right>(\varepsilon)$ is obtained by averaging 
$S_{av}$ over a large number of samples at the same deformation $\varepsilon$.
It can be observed in Fig.\ \ref{fig:clust_num}$(a)$ that in the brittle 
regime $\alpha < \alpha_c$ the average cluster size
$\left<S_{av}\right>(\varepsilon)$ has only an increasing branch which suddenly stops
at macroscopic failure.
However, in the ductile phase 
the $\left<S_{av}\right>$ curves have a maximum
followed by a rapidly decreasing part. Above the critical point $\alpha>\alpha_c$
the coalescence of cracks results in a macroscopic 
cluster which spans the entire system. 
The maximum of the $\left<S_{av}\right>$ curves occurring at the deformation $\varepsilon_m^{S}$ 
marks the configuration where the dominating
cluster is formed. Since the largest cluster is always removed from
the statistics, in the presence of a dominating cluster $\left<S_{av}\right>$ must 
decrease. The situation drastically changes when $\alpha$ surpasses the value of 
$\alpha^*=1-p_c\approx 0.4077$ because in the regime $\alpha>\alpha^*$ the weak
component does not develop a spanning cluster and hence no dominating crack can
emerge. Consequently, in this regime $\left<S_{av}\right>$ monotonically increases
and saturates when nearly all weak fibers break.

\begin{figure}
\begin{center}
\epsfig{bbllx=10,bblly=495,bburx=350,bbury=775,file=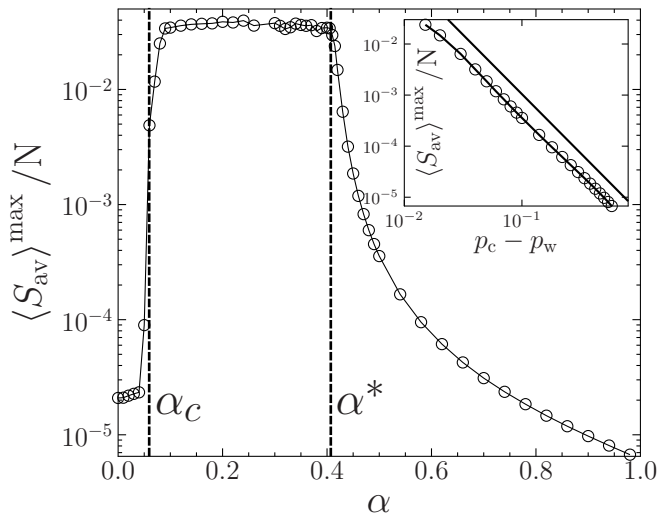,
  width=8.5cm}
  \end{center}
   \caption{The maximum value of the average cluster size
   $\left<S_{av}\right>^{max}$ averaged over a large number of samples at a given
   $\alpha$. Three distinct regimes can clearly be distinguished which are
separated by $\alpha_c$ and $\alpha^*$. The inset presents the same quantity in the range
$\alpha > \alpha^*$ as a function of $p_c-p_w$, where $p_w=1-\alpha$. The straight line has
slope $2.32$.
\label{fig:clust_average_alpha}}
\end{figure}

It can be observed in Fig.\ \ref{fig:clust_num}$(a)$ that the value of the maximum  
$\left<S_{av}\right>^{max}$ of the average cluster size shows also 
an interesting systematics as the composition $\alpha$ of the system is changed. 
For clarity, we also present separately the maximum 
$\left<S_{av}\right>^{max}$ as a function of $\alpha$ in Fig.\ \ref{fig:clust_average_alpha}.
In the brittle phase $\alpha<\alpha_c$ the maximum $\left<S_{av}\right>^{max}$ is 
simply the value of the terminal point of the function $\left<S_{av}\right>(\varepsilon)$ at 
the instant of failure which is practically independent of $\alpha$. However,
as $\alpha$ surpasses the critical point $\alpha_c$, the coalescence of clusters results in crack sizes
which are orders of magnitude larger than  the ones observed in the brittle regime. Inside the ductile regime
$\alpha>\alpha_c$ the maximum remains constant since it is only determined by the system size
$N$ until the weak fibers have a spanning cluster $\alpha_c\leq\alpha\leq\alpha^*$.
However, for $\alpha>\alpha^*$ the maximum of $\left<S_{av}\right>(\varepsilon)$  
is reached at saturation, when all the weak clusters have been broken. 
Hence, in this regime the maximum value $\left<S_{av}\right>^{max}$ is a decreasing
function of $\alpha$ in Fig.\ \ref{fig:clust_average_alpha}. It is interesting
to note that approaching $\alpha^*$ from above $\alpha\to\alpha^*_+$ the occupation
probability of the weak component $p_w=1-\alpha$ tends to $p_c$ from below.
Consequently, in the regime $\alpha>\alpha^*$, the maximum value of the average crack size 
$\left<S_{av}\right>^{max}$  has the same critical behavior as the average cluster size 
of percolation when approaching the  critical point
\begin{eqnarray}
\left<S_{av}\right>^{max} \sim (p_c-p_w)^{-\gamma}.
\end{eqnarray}
The inset of Fig.\ \ref{fig:clust_average_alpha} demonstrates that the above prediction
is perfectly fulfilled by the simulation data.
The numerical value of the exponent $\gamma=2.33\pm 0.08$ falls very close to the corresponding 
exponent of percolation $\gamma=43/18\approx 2.39$ \cite{stauffer_introduction_1992}.

The average number of cracks $\left<n_c\right>$ also encodes interesting
information about the evolution of the crack ensemble. 
It can be observed in Fig.\ \ref{fig:clust_num}$(b)$ that
in the brittle regime $\alpha<\alpha_c$ the number of cracks monotonically
increases and stops suddenly due to the abrupt failure of the system 
(compare to Fig.\ \ref{fig:constitutive}).
When the fracture is ductile $\alpha\geq\alpha_c$ the $\left<n_c\right>$ curves
develop a decreasing regime due to the coalescence of cracks such that 
the position of the maximum $\varepsilon_m^{nc}$ marks the configuration where coalescence
starts. For clarity, the maximum position $\varepsilon_m^{nc}$ is indicated by the vertical lines in Fig.\
\ref{fig:clust_num} for a few $\alpha$ values. Comparing Fig.\ \ref{fig:clust_num}$(a)$ 
and Fig.\ \ref{fig:clust_num}$(b)$ it can be observed that when the strain $\varepsilon$ 
exceeds $\varepsilon_m^{nc}$ the average cluster size displays a faster growth  
and reaches its maximum at a larger strain value $\varepsilon_m^{S}$.
For $\alpha<\alpha_c$ these two deformations coincide at 
$\varepsilon_m^{nc} = \varepsilon_m^{S} = \varepsilon_c^{\sc LLS}$, where macroscopic failure takes place,
however, in the ductile regime the relation $\varepsilon_m^{nc} < \varepsilon_m^S$ holds.
Only cracks that develop inside larger clusters of weak fibers can merge, therefore, isolated weak
clusters break at higher loads due to the global load reduction associated to the fraction 
of the load sustained by strong fibers. This morphological feature explains 
the increase of $\left<n_c\right>$ at large strains, as displayed in Fig.\ \ref{fig:clust_num}$(b)$.
It is important to emphasize that the decreasing branch of $\left<n_c\right>$
prevails even above $\alpha^*$, which shows that merging can occur not only inside
the spanning cluster, but
even smaller weak clusters may develop several cracks which later 
on merge. 

The most remarkable result, presented in Fig.\ \ref{fig:clust_num}$(b)$, 
is that the maximum value of the cluster number remains constant 
$\left<n_c\right>/N\approx 0.1$ until the strong
component does not have a spanning cluster $\alpha<p_c$. In this parameter regime 
the maximum cluster number is determined solely by the distribution of the breaking 
thresholds of the weak fibers. Our result indicates that for 
a uniform distribution of breaking thresholds, approximately, $10\%$ of the lattice 
sites serve as nucleation centers of
cracks. It is noticeable that until $\alpha<p_c$ the presence of strong fibers only shifts
the value of the deformation $\varepsilon_m^{nc}$ where the maximum is reached.
The disappearance of merging is indicated by the monotonicity of $\left<n_c\right>(\varepsilon)$
which occurs when $\alpha$ exceeds $p_c$. 
In this regime, strong fibers form a spanning cluster and weak clusters are 
so small that they only nucleate one crack, which gradually grows covering 
the entire cluster.

In order to quantify the effect of strong fibers on the evolution of the number of cracks, 
Fig.\ \ref{fig:numcollap_csize}$(a)$ displays $\left<n_c\right>/N$ as a function 
of $\varepsilon (1-\alpha)$. 
A high quality data collapse is obtained up to the maximum, which implies 
that up to deformations of order
\begin{eqnarray}
 \varepsilon_m^{nc} \sim \frac{1}{1-\alpha},
\end{eqnarray}
the crack nucleation dominates the behavior of damaged domains, 
while above $\varepsilon_m^{nc}$  crack merging controls the evolution of 
the micro-structure of the system. 
This behavior implies that  large bursts of breaking fibers become relevant 
for $\varepsilon>\varepsilon_m^{nc}$ and they are mainly growth steps of existing cracks.  
Above the percolation threshold ($\alpha > p_c$), strong fibers prevent the merging
of cracks nucleated within isolated weak clusters. In this case
cracks must nucleate even for $\varepsilon>\varepsilon_m^{nc}$ to break all weak
fibers, and hence, $\left<n_c \right>$ becomes monotonically increasing.

\subsection{Size distribution of cracks} 
The statistics of crack sizes plays an important role in the emergence of localization 
and macroscopic failure of heterogeneous materials under an increasing external load
\cite{zapperi_first-order_1997,kun_damage_2000-1,PhysRevE.65.056105,PhysRevE.73.036109,nanjo_damage_2005}. 
In the brittle phase $\alpha<\alpha_c$ of our FBM we evaluated the size distribution $P(S)$
of cracks in the last stable configuration of the bundle before the catastrophic 
avalanche.  It has been discussed above that in the ductile phase,  $\alpha>\alpha_c$,
the onset of localization is marked by the position of the maximum $\varepsilon_m^{nc}$
of the number of cracks. However, the statistics and structure of cracks 
at this deformation is rather close to the limiting case of the absence of strong fibers  
$\alpha=0$. That's why the size distribution of cracks $P(S)$ 
was evaluated at the deformation $\varepsilon_m^S$ where the average cluster size 
$\left<S_{av}\right>(\varepsilon)$ has its maximum. 

\begin{figure}
\begin{center}
\epsfig{bbllx=5,bblly=450,bburx=700,bbury=750,file=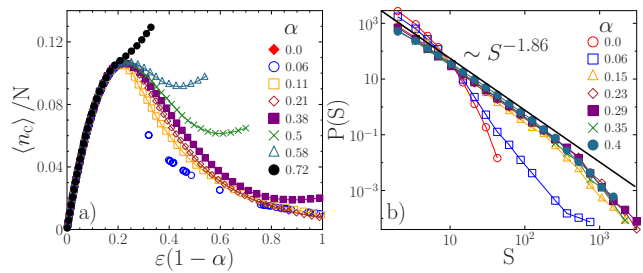,
  width=8.5cm}
   \caption{{\it (Color online)} $(a)$ Average number of cracks as a function of 
$\varepsilon$ rescaled along the horizontal axis with $p_w=1-\alpha$. 
High quality data collapse can be observed
up to the maximum. $(b)$ Cluster size distributions determined at $\varepsilon_m^{S}$
for several compositions $\alpha$.}
\label{fig:numcollap_csize}
  \end{center}
\end{figure}
\begin{figure}
\begin{center}
\epsfig{bbllx=0,bblly=0,bburx=590,bbury=590,file=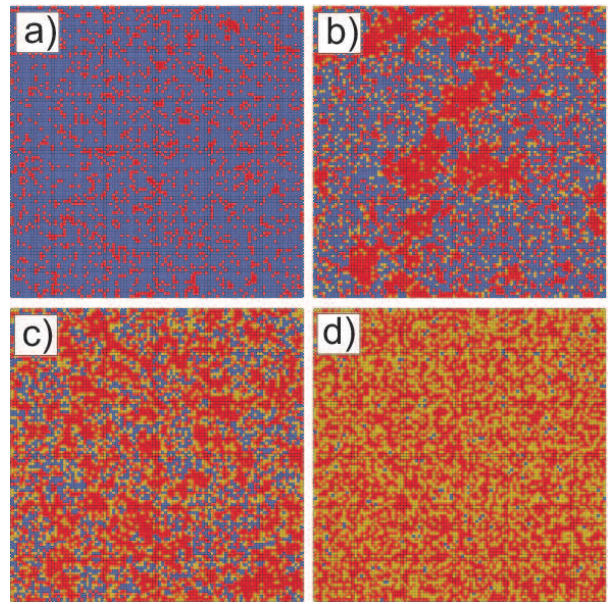,
  width=8.0cm}
  \end{center}
   \caption{{\it (Color online)}
  Snapshots of the cluster structure in systems of size $L=100$
taken at the deformation $\varepsilon_m^{S}$ where the average cluster size $\left<S_{av}\right>$ has
a maximum for different values of the control parameter $\alpha$: $(a)$ $0$ $(b)$
$0.08$ $(c)$ $0.35$ $(d)$ $0.4$. The assignment of colors is the same as in Fig.\ 
\ref{fig:continous_state_matrix}.
When the system only contains weak fibers
$(a)$ all the clusters are small compared to the system size. Above the critical point
$\alpha_c$ growth and merging result in large cluster sizes which are only limited
by the system size and by the underlying lattice structure of the weak and strong components.
\label{fig:final_state_matrix}}
\end{figure}

Figure \ref{fig:final_state_matrix} presents the configuration of the bundle at $\varepsilon_m^S$ for several
values of $\alpha$. 
It can be observed in Fig.\ \ref{fig:final_state_matrix}$(a)$ that if the bundle
only contains weak fibers $\alpha=0$ all clusters are small compared to the system size.
Consequently, in Fig.\ \ref{fig:numcollap_csize}$(b)$ the corresponding size distribution $P(S)$ 
is a rapidly decreasing exponential in agreement with former results on LLS FBMs 
\cite{zapperi_first-order_1997,kun_damage_2000-1,hidalgo_fracture_2002}. 
In the ductile regime $\alpha>\alpha_c$
the spanning cluster covers a large fraction of the system (see Fig.\ 
\ref{fig:final_state_matrix}$(b,c,d)$), however, the size of smaller cracks scatter
over a broad interval, as well. It is important to note that in these cases the cluster size
distribution has a power law regime
\begin{eqnarray}
P(S) \sim S^{-\tau},
\end{eqnarray}
followed by an exponential cutoff. The value of the exponent $\tau=1.86\pm 0.05$ 
proved to be lower than the corresponding exponent of percolation $\tau=187/91\approx 2.05$ 
\cite{stauffer_introduction_1992}. It shows that larger cracks
more frequently occur due to the correlated growth mechanism and coalescence.
It is interesting to note that our results on the size distribution of cracks
have a good qualitative agreement with Ref.\ \cite{PhysRevE.65.056105}, where a two dimensional
fracture model predicted power law distributed crack sizes with exponent $\tau = 2$. The authors 
argued that the growth and coalescence of cracks along the softening regime of the
constitutive curve are responsible for the scale free distribution. However,
when the failure is brittle or quasi-brittle 
exponential or lognormal distributions have been obtained in various types of fracture
models \cite{zapperi_first-order_1997,kun_damage_2000-1,PhysRevE.73.036109,PhysRevE.86.025101}
again in agreement with our results.

\section{Discussion}
We have examined the fracture process of highly heterogeneous materials in the 
framework of a fiber bundle model. The fibers have identical elastic properties,
however, their fracture characteristics are strongly different: a fraction of 
fibers is strong in the sense that they can sustain any load without breaking,
while the rest of fibers is weak having statistically distributed strength values.
The two components are homogeneously mixed on a square lattice which is then loaded
in a stress controlled way.
In order to capture the effect of stress concentration around cracks, we considered
localized load sharing such that the load of broken fibers is redistributed 
equally over its intact nearest neighbors on the lattice.
Our computer simulations revealed that varying the fraction of strong fibers $\alpha$ 
as a control parameter the mechanical heterogeneity leads 
to a rich  variety of mechanical responses.

Investigating the macroscopic constitutive behavior of the system we pointed out that
at a critical value $\alpha_c$ a transition occurs from highly brittle
to ductile response. The brittle phase is characterized by the presence of an
unstable branch of the constitutive curve along which a macroscopic fraction 
of fibers breaks in an abrupt avalanche
when performing stress controlled loading.
In the ductile regime the constitutive curve is monotonous
with a relatively long plateau regime, which implies that weak 
fibers break gradually through 
finite size avalanches leading to stable fracture.
The critical fraction, $\alpha_c$, has an astonishingly small value showing
that adding a very small amount of strong fibers can be sufficient to stabilize 
an originally brittle system.

On the microscopic scale our simulations revealed that the interplay of the
inhomogeneous stress field arising due to the localized load redistribution
and of the load-absorbing effect of strong fibers leads to a rich dynamics.
In the ductile regime 
the size distribution of avalanches proved to be a power law with an exponent 
$\mu=2.0$, which is significantly lower than the one of the brittle phase $\mu=9/2$.
Approaching the critical fraction 
from above, the characteristic  avalanche size has a power law divergence
which shows that
the brittle-ductile transition occurs analogous to continuous phase transitions.

The localized load redistribution has the consequence that avalanches of fiber 
breakings form spatially connected clusters. Since high stress is concentrated
along the perimeter of broken clusters, avalanches are typically triggered at these
fibers so that avalanches are intermittent steps of the growth of cracks.
Our investigations showed that in the brittle regime only small cracks can develop
compared to the system size. However, in the ductile phase the load absorbed
by the strong fibers allows the cracks to reach sizes where they merge
and form a macroscopic crack spanning the entire system. Merging always occurs
until it becomes prevented by the spanning cluster of the strong component in
the limit of large values of the control parameter $\alpha$.

Analyzing the statistics of bursts and cracks, we pointed out that at the 
beginning of the loading process the nucleation of cracks dominates, while later on
coalescence governs the evolution of the micro-structure. Since the maximum number of cracks
does not change, bursts emerging along the plateau of the constitutive curve in the
ductile regime are steps of stable crack propagation. 

\begin{acknowledgments}
The work is supported by TAMOP-4.2.2.A-11/1/KONV-2012-0036, 
TAMOP-4.2.2/B-10/1-2010-0024, and OTKA K84157 projects. The
project is implemented through the New Hungary Development Plan,
co-financed by the European Union, the European Social Fund and 
the European Regional Development Fund.
This work was supported by the European Commissions by the
Complexity-NET pilot project LOCAT and by NFÜ under the contract ERANET\_HU\_09-1-2011-0002. 
I.P. acknowledges financial
support from MICINN (Spain) under projects $FIS2008-06034-C02-02$, $FIS2011-22603$ and DURSI $SGR2009-00634$.
\end{acknowledgments}

\bibliography{../pap}
\end{document}